\documentclass[twocolumn,floatfix,pre,showpacs]{revtex4-1}
\pdfoutput=1
\usepackage{graphicx}
\usepackage{amssymb}
\usepackage{amsmath}
\usepackage{colonequals}
\usepackage[utf8]{inputenc}
\usepackage{soul}
\usepackage{cancel}
\newcommand{\kBT}{k_\mathrm{B}T}

\newcommand{\FF}{\mathcal{F}}
\newcommand{\tauB}{\tau_\mathrm{B}}

\begin{document}
\title{Enhancement of mobility in an interacting colloidal system under feedback control}
\author{Robert Gernert and Sabine H.~L.~Klapp}
\affiliation{
Institut f\"ur Theoretische Physik, Sekr.~EW 7--1,
Technische Universit\"at Berlin, Hardenbergstrasse 36,
D-10623 Berlin, Germany
}
\date{\today}

\begin{abstract}
Feedback control schemes are a promising way to manipulate transport properties of driven
colloidal suspensions. In the present article we suggest a feedback scheme to enhance the 
{\em collective} transport of colloidal particles with repulsive interactions through a one-dimensional tilted washboard potential. The control is modelled by a 
harmonic confining potential, mimicking an optical ``trap'', with the center of this trap moving with the (instantaneous) mean particle position. Our
theoretical analysis is based on the Smoluchowski equation combined with Dynamical Density Functional Theory (DDFT) 
for systems with hard-core or ultra-soft (Gaussian) interactions. 
For either type of interaction we find that the feedback control
can lead to an enhancement of the
mobility by several orders of magnitude relative to the uncontrolled case. The largest effects occur for intermediate stiffness of the trap and large particle numbers. Moreover, in some regions of the parameter space the feedback control induces oscillations of the mean velocity. Finally, we show that the enhancement of mobility is robust against a small time delay in implementing the feedback control.
\end{abstract}
\maketitle

\section{Introduction}
The manipulation of colloidal transport properties with feedback mechanisms has become an active topic of research in recent years. Examples include 
 the improvement of the net current in one-dimensional ratchet systems \cite{lopez08,toyabe10,craig08-024,loos14}, the transport of interacting particles in a tilted washboard under Pyragas control \cite{lichtner12-300,lichtner10}, the sorting of colloids in a micro-fluidic channel \cite{prohm14}, and the
adjustment of viscosity of a sheared colloidal suspension \cite{vezirov15}. Further, feedback control has become an important concept in (bio-)particle trapping \cite{balijepalli12,qian13-370,bregulla14,braun14,jun14,cole12}, reaction-diffusion systems \cite{loeber14-431}, quantum transport \cite{emary13,brandes10,poeltl11}, laser dynamics \cite{dahms10}, and brain dynamics \cite{masoller08,lehnert11}. An essential factor supporting the development of and theoretical research on feedback control strategies
is the recent major progress of corresponding experimental techniques, among them the monitoring and steering of colloids \cite{qian13-370,braun14,selmke14} and biomolecules \cite{fisher05}, and preparation and non-destructive (weak) measurement of quantum states \cite{sayrin11}.

Within colloidal transport, most of the feedback studies so far involve single particles \cite{bregulla14,abreu11,bauer12} or dilute suspensions \cite{lopez08,craig08-024}, i.e., 
systems of non-interacting particles. We note that, even in this idealized situation, feedback can induce {\em effective} interactions if the protocol involves system-averaged quantities \cite{cao04}. For many real colloidal systems, however, direct interactions between the colloids stemming e.g., from excluded volume effects, charges on the particles' surfaces, or (solvent-induced) depletion effects cannot be neglected. First studies of feedback control in presence of colloidal interactions indicate indeed complex dynamical scenarios. An example
was considered in Refs. \cite{lichtner12-300,lichtner10}, where a Pyragas-type control of colloidal transport in one dimension resulted in current reversal and oscillatory states.

In the present work we explore the transport of interacting (repulsive) colloids in presence of a feedback-controlled harmonic ``trap". 
Indeed, trap-like devices appear as a standard tool to implement feedback, both in experiments (see, e.g., \cite{lopez08,qian13-370,braun14,balijepalli12,cole12,mirowski05}) and in theory
\cite{abreu11}. A prominent example is an optical laser tweezer acting on polarizable colloids. The corresponding trap potential  
can be modelled as a quadratic function in space \cite{cole12,florin98,loewen01}.

In conventional applications the position of the center of the trap acting on the colloidal particle(s) is either constant in space, 
or it moves in an externally prescribed manner \cite{lee04,blickle06,florin98}. In contrast to these situations (which are termed ``open-loop" in control theory), we here
consider a harmonic trap whose center coincides with the {\it mean} position of the particles. Thus, the trap potential depends on the particle's position, yielding a feedback scheme.

As a model system to demonstrate the principle of this feedback control we consider the paradigm example of colloids driven through a one-dimensional, spatially oscillating, tilted ``washboard" potential with energy barriers much larger than the thermal energy. 
Already without feedback or any trap potential these (overdamped) systems show interesting effects such as absolute negative mobility \cite{speer12-179} or enhancement of diffusion at  a certain ``critical" force \cite{evstigneev09,herrera07}. Many single-particle transport properties in washboard potentials can be derived analytically \cite{risken84,reimann01-035,sancho10,gernert14}. Further,
recent numerical studies indicate interesting interaction-induced transport phenomena,
examples being coherent motion of attractively interacting particles \cite{pototsky10,evstigneev09,speer12-179}, density excitations in Frenkel-Kontorova models \cite{noguera13}, or single file diffusion \cite{herrera07}. Given this background,
one may expect that the interplay of external potentials, particle interactions, and feedback yields exciting additional effects. Our study shows that this is indeed the case.

Specifically, we consider particles with either (infinitely) hard or soft (Gaussian) repulsion, the latter describing polymeric particles in a coarse-grained fashion 
\cite{stillinger76,likos01}. The feedback control is implemented on the level of the Smoluchowski (overdamped Fokker-Planck) equation, in which the particle interactions are treated
via Dynamical density functional theory (DDFT) \cite{archer04}. Our numerical results demonstrate that the 
feedback-controlled trap in conjunction with particle interactions can lead to a drastic increase of the mobility by orders of magnitude.
Loosely speaking,  the particles ``help to push each other over the energy barrier''. This phenomenon is accompanied by 
a freezing of the width of the density distribution (thus opposing normal diffusion), and to
 time-periodic oscillations of the mean velocity not seen in the uncontrolled case.
 
In the major part of our study we assume instantaneous feedback. This is clearly an idealization
given the fact that, in an experiment with feedback control, there is always one (or several) time delay(s) due to measurement, information processing, and implementation of the forcing \cite{craig08-158}.
However, the time delay of modern experimental feedback techniques for colloids \cite{lopez08,qian13-370,cohen06,jun14} is much smaller than the time scale of particle motion, justifying the approximation of instantaneous feedback. Still, to estimate the effects we also consider briefly the impact of time delay.

The remainder of this paper is organized as follows.
After the introduction of the theoretical background in Sec.~\ref{sec:model}, we discuss in Sec.~\ref{sec:sp} the effect of feedback on a single particle. In the limit
of vanishing washboard potential the transport can here be calculated analytically. The full problem is discussed in Sec.~\ref{sec:int}, where we present the main results. A conclusion is given in Sec.~\ref{sec:conclusion}.
\section{Model}\label{sec:model}
We consider the motion of $N$ interacting Brownian particles in one dimension under the influence of the externally imposed, tilted washboard potential
\begin{align}
	V_\mathrm{ext}(x)&=u(x)-x\,F
	\label{Vext}
	\,,
\end{align}
where $u(x)=u_0\sin^2(\pi x/a)$. In Eq.~\eqref{Vext},
$F$ denotes a constant driving force, and $a$ and $u_0$ denote wavelength and amplitude of the periodic potential $u(x)$, respectively.
On the particle level, the motion is described by the $N$ coupled, overdamped Langevin equations
\begin{align}
	\gamma\dot{x}_i(t)=&-u'(x_i)+F+f_i^\mathrm{int}(x_1,\dots,x_N)
	\notag
	\\
	&+f_i^\mathrm{DF}(x_1,\dots,x_N)+\sqrt{2\kBT\gamma}\,\xi_i(t)
	\label{LE}
\end{align}
for the position $x_i(t)$ of the $i$th particle \cite{risken84}. In Eq.~\eqref{LE}, the friction constant is denoted by $\gamma$, Boltzmann's constant by $k_\mathrm{B}$, the temperature by $T$, and the $\xi_i(t)$ are independent random numbers chosen from a Gaussian distribution with zero mean and unit variance. Further, $f_i^\mathrm{int}$ represents the force due to interaction between particle $i$ and other particles $j$, that is $f_i^\mathrm{int}=-\partial/\partial x_i \sum_{j\neq i} v(x_i,x_j)$ and $f_i^\mathrm{DF}$ is the force due to feedback control, where $\mathrm{DF}$ stands for \emph{dynamic freezing}. These forces are specified below.
\par%
In the present study we rather describe the motion in terms of the space- and time-dependent one-particle density \mbox{\cite{hansen06,archer04}}
\begin{align}
	\varrho(x,t)&=\left\langle\sum_{i=1}^N\delta(x-x_i(t))\right\rangle
	\label{rho}
	\,,
\end{align}
where $\langle\dots\rangle$ denotes an average over all realisations of the random force $\xi_i(t)$. The density is normalized according to $\int\mathrm{d}x\,\varrho(x,t)=N$. The time evolution of $\varrho(x,t)$ is governed by the extended Smoluchowski equation
\begin{align}
	\partial_t\varrho(x,t)=&\frac{\kBT}{\gamma}\partial_{xx}\varrho(x,t)+\frac{1}{\gamma}\partial_x\left(\varrho(x,t)\partial_x[V_\mathrm{ext}(x)\right.
	\notag
	\\
	&\left.+V_\mathrm{DF}(x,\varrho)+V_\mathrm{int}(x,\varrho)]\right)
	\label{ddft}
	\,,
\end{align}
where the impact of particle interactions and of feedback control enters via the potentials $V_\mathrm{int}$ and $V_\mathrm{DF}$, respectively. Specifically, to treat the particle interactions we employ the concepts of Dynamical Density Functional Theory (DDFT) \cite{marconi99,archer04,espanol09}. 
In this framework,
\begin{align}
	V_\mathrm{int}(x,\varrho)&=\frac{\delta \FF_\mathrm{int}[\varrho]}{\delta\varrho(x,t)}
	\,,
	\label{vint}
\end{align}
where $\FF_\mathrm{int}[\varrho]$ is the interaction part of an \textit{equilibrium} free energy functional and $\delta/\delta\varrho$ denotes a functional derivative.
Equation \eqref{vint} implicitly contains an adiabatic approximation, i.e., the assumption that non-equilibrium correlations, at each time $t$, can be replaced by those of an equilibrium system with density $\varrho(x,t)$.
\par%
We consider two types of interacting systems, that is, ultra-soft particles described by the Gaussian core model (GCM) and hard particles.
The pair interaction potential according to the Gaussian core model reads
\begin{gather}
	v_\mathrm{GCM}(x_i,x_j)=\varepsilon\exp\left(-\frac{(x_i-x_j)^2}{\sigma^2}\right)
	\label{vgcm}
	\,.
\end{gather}
This potential has been introduced as a coarse-grained (center-of-mass) potential between two fluctuating polymer chains, with the particle diameter $\sigma$ being proportional to the polymers' radius of gyration \cite{stillinger76,likos01,goetze06}.
To incorporate the GCM interactions into the dynamical equation \eqref{ddft}, we employ the mean-field free energy functional
\begin{gather}
	\FF_\mathrm{int}^\mathrm{GCM}[\varrho]=\frac{1}{2}\!\int\!\mathrm{d}x\,\mathrm{d}x'\,\varrho(x,t)\,v_\mathrm{GCM}(x-x')\,\varrho(x',t)
	\label{Fgcm}
	\,.
\end{gather}
This functional has been proven to give a reliable description of the equilibrium structure of the GCM, particularly at intermediate and high densities \mbox{\cite{likos14}}. Combining Eq.~\eqref{Fgcm} and \eqref{vint}, we obtain
\begin{gather}
	V_\mathrm{int}^\mathrm{GCM}(x,\varrho)=\int\mathrm{d}x'\,\varrho(x',t)\,v_\mathrm{GCM}(x-x')
	\label{Fmf}
	\,.
\end{gather}
\par%
Hard particles with diameter $\sigma$ are described by the interaction potential
\begin{gather}
	v_\mathrm{hard}(x_i,x_j)=\begin{cases} 0 &,\;\text{for}\;\vert x_i-x_j\vert \ge \sigma \\\infty&,\;\text{for}\;\vert x_i-x_j\vert<\sigma \end{cases}
	\label{vhc}
	\,.
\end{gather}
For one-dimensional systems of hard spheres there exists an exact free energy functional \cite{percus76}
\begin{align}
	\FF_\mathrm{int}^\mathrm{hard}[\varrho]=-\frac{1}{2}\int\mathrm{d}x&\,\ln\left(1-\int_{x-\sigma/2}^{x+\sigma/2}\mathrm{d}x'\,\varrho(x',t)\right)
	\notag
	\\
	&\times\![\varrho(x+\frac{\sigma}{2},t)+\varrho(x-\frac{\sigma}{2},t)]
	\label{Fhc}
	\,,
\end{align}
The free energy \eqref{Fhc} corresponds to the one-dimensional limit of fundamental measure theory \cite{roth10}.
\par%
We now turn to the modelling of feedback control. To this end we use the potential
\begin{gather}
	V_\mathrm{DF}(x,\varrho)=\eta (x-\langle x\rangle)^2
	\label{Vdf}
	\,,
\end{gather}
where
\begin{gather}
	\langle x\rangle(t)=\frac{1}{N}\int\mathrm{d}x\,x\,\varrho(x,t)
	\label{mean}
\end{gather}
is the time-dependent mean particle position. Thus, Eq.~\eqref{Vdf} describes a moving harmonic trap centered around the mean position, resembling the potential seen by particles in moving optical traps \mbox{\cite{florin98,cole12}}. The strength of the harmonic confinement, $\eta$, is set to constant.
Since $V_\mathrm{DF}$ depends on $\langle x\rangle(t)$ and, thus, on the dynamical state of the system, it corresponds to a true feedback control. This is different from an ``open-loop controlled'' moving trap, where $\langle x\rangle$ in Eq.~\eqref{Vdf} would be replaced by a position moving with fixed velocity $v_0$.
\par%
We also note that the fact that our feedback control is coupled to a ensemble averaged quantity is in contrast to other feedback mechanisms which are based on individual particle positions \mbox{\cite{craig08-158}}.
At the level of the Langevin equation \eqref{LE} our feedback control force reads $f_i^\mathrm{DF}(x_i,\langle x\rangle)=2\eta (x_i-\langle x\rangle(t))$. Thus $f_i^\mathrm{DF}$ only depends on a single coordinate, $x_i$, and on $\langle x\rangle$. This differs from other feedback control approaches where the feedback force itself depends on the number of particles \mbox{\cite{craig08-158,cao04,craig08-024}}.
\section{Single-particle transport}\label{sec:sp}
To understand the basic properties of the effect of the feedback control Eq.~\eqref{Vdf} we first discuss the single-particle case ($N\!=\!1, \FF_\mathrm{int}=0$) without the periodic potential ($u_0\!=\!0$). In this case Eq.~\eqref{ddft} reduces to the one-dimensional Smoluchowski equation
\begin{gather}
	\gamma\partial_t\varrho=\kBT\partial_{xx}\varrho+\partial_x(\varrho(V_\mathrm{DF}'- F))
	\label{se-simplest}
	\,.
\end{gather}
A main quantity characterizing the transport is the mean particle position $\langle x\rangle$, defined in Eq.~\mbox{\eqref{mean}}, as function of time. Solving Eq.~\eqref{se-simplest} analytically with the initial condition $\varrho(x,t\!=\!0)=\delta(x-x_0)$ yields
\begin{align}
	\langle x\rangle(t)&=\frac{F}{\gamma}\,t\,+x_0
	\label{mean-free-diff}
	\,.
\end{align}
Equation \eqref{mean-free-diff} shows that the mean particle position does not depend on the confinement strength $\eta$. This can also be seen by applying the coordinate transformation $x'=x-vt$ to the Smoluchowski equation Eq.~\eqref{se-simplest}, setting $v=F/\gamma$. With this transformation the term $\partial_x(\varrho F)$ vanishes. Further, the force related to $V_\mathrm{DF}$ is invariant with respect to this transformation. Hence, the influences of $F$ and $\eta$ decouple. From Eq.~\eqref{mean-free-diff} we calculate the mobility
\begin{align}
	\mu&\colonequals \lim_{t\to\infty}\frac{\partial_t \langle x\rangle}{F}
	\label{mu}
	\\
	&=\frac{1}{\gamma}
	\label{mobilityfree}
	\,,
\end{align}
which only depends on the friction constant $\gamma$. We will refer to this value of $\mu$ as the mobility of free motion.
\par%
A further quantity of interest is the mean squared displacement
\begin{align}
	w(t)&=\langle (x-\langle x\rangle)^2\rangle
	\label{width}
	\,.
\end{align}
To calculate $w(t)$ we use $\langle V_\mathrm{DF}\rangle=\eta w$ [cf.~Eqs.~\mbox{\eqref{Vdf}} and \mbox{\eqref{se-simplest}}] which yields
\begin{align}
	w(t)&=\frac{\kBT}{2\eta}\left(1-e^{-4\eta t/\gamma}\right)
	\label{width-free-diff}
	\,.
\end{align}
For short times $w(t)$ growths linearly with time, corresponding to diffusive behaviour. For long times diffusion is suppressed: $w(t)$ approaches a limiting value determined by $\eta$.
Interestingly, a similar behaviour of $w(t)$ occurs in a model of feedback control of quantum transport \cite{brandes10}.
There, the fluctuations of the number of electrons tunneling through a quantum junction are suppressed with a feedback control force, which is linear in the fluctuation of the number of electrons. This corresponds to our harmonic confinement of the density fluctuation, and indeed, the two physically different situations are describable by a formally identical Smoluchowski equation \cite{strasberg14}.
\par%
We now turn to the system in presence of the potential $u(x)$, defined below Eq.~\eqref{Vext}.
In the single particle case ($N\!=\!1, \FF_\mathrm{int}\!=\!0$) Eq.~\mbox{\eqref{ddft}} then reduces to the Smoluchowski equation
\begin{gather}
	\gamma\partial_t \varrho=\kBT\partial_{xx}\varrho+\partial_x(\varrho(V'_\mathrm{DF}+V'_\mathrm{ext}))
	\label{se}
	\,.
\end{gather}
Without control ($\eta\!=\!0$) Eq.~\mbox{\eqref{se}} describes the thoroughly studied case of a Brownian particle in a washboard potential, where the mobility, as well as the long-time diffusion constant, are accessible analytically \cite{stratonovich58,reimann01-035,risken84}.
From that it is known that the mobility is very small if $u_0 \gg \kBT$ and if the driving force $F$ is smaller than the so-called critical force $F_c=u_0\pi/a$, related to the diffusion maximum \mbox{\cite{reimann01-035}} (for $F>F_c$ the potential minima vanish).
Otherwise the mobility is large, in particular it approaches $1/\gamma$ for $u_0/F\to 0$. The goal of our study is to enhance the mobility in the regime of deep wells. 
\subsection{Numerical Results}\label{sec:sp-results}
To explore the single particle transport for finite $\eta$ and $V_\mathrm{ext}\neq 0$ we solve Eq.~\mbox{\eqref{se}} numerically, choosing $u_0=15\kBT$ and $F=0.2 F_c$. As initial condition we choose the equilibrium (Boltzmann) distribution corresponding to the case $F=0$
\begin{align}
	\varrho(x,0)&=\exp\left(-(\eta\,x^2+u(x))/\kBT\right)/Z
	\label{sp-ic}
	\,,
\end{align}
where $Z$ is a normalisation constant.
\par
\begin{figure}%
	\centering
	\includegraphics[width=\linewidth]{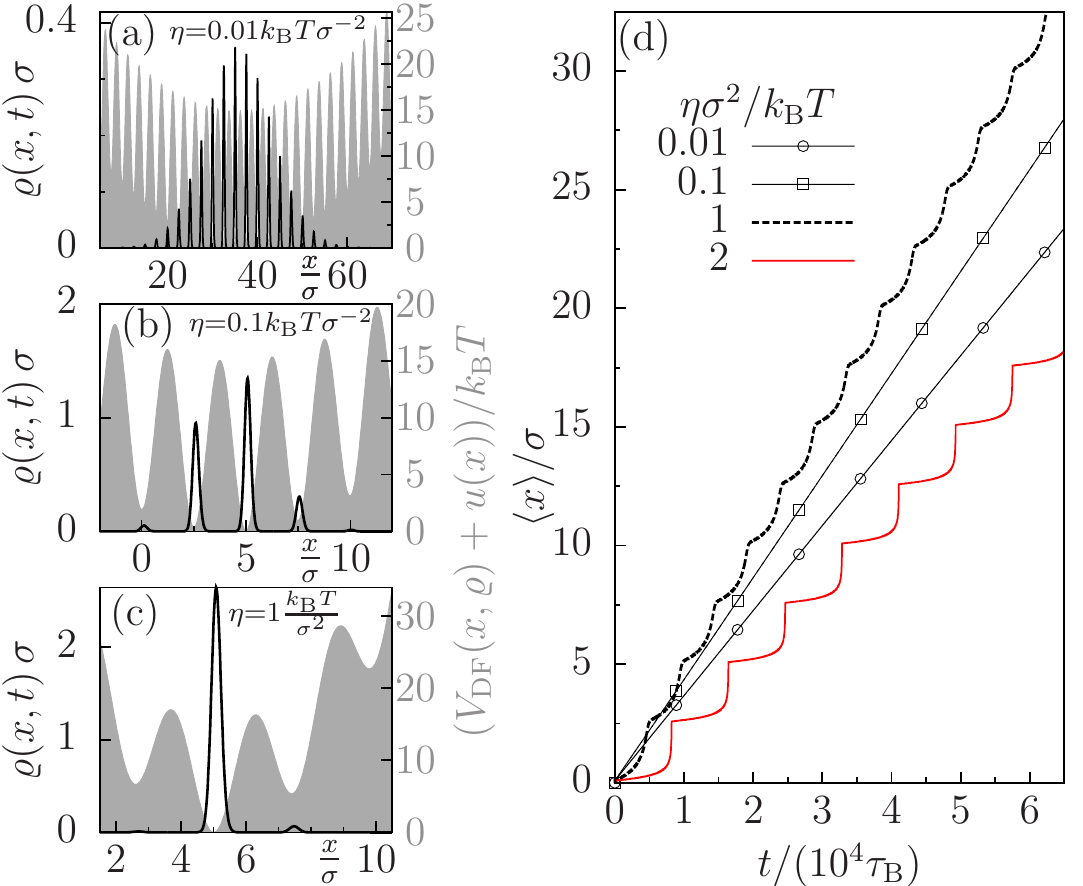}
	\caption{Single particle transport through a potential with wavelength $a=2.5\sigma$. (a-c) Density plots (black, left axis) and potential $V_\mathrm{DF}+u$ (grey, right axis). The time $t$ is (a) $10^5\tauB$ and (b,c) $10^4\tau_B$ where the Brownian time $\tauB=\sigma^2\gamma/(\kBT)$. Part (d) shows the mean particle position with respect to time for different $\eta$. For strong confinements an oscillatory behaviour emerges.}
	\label{rho-mean}
\end{figure}

%
Figures \ref{rho-mean}(a-c) show plots of the one-particle density $\varrho(x,t)$ for three values of $\eta$.
As expected for a trap, the width of the density distribution becomes the smaller the larger $\eta$.
Figure \mbox{\ref{rho-mean}(d)} shows additionally the mean particle position with respect to time.
Interestingly, we find that at large values of $\eta$, oscillatory solutions emerge.
At the corresponding values of $\eta$ the confinement is so strong that the particle is confined to a single well of the periodic potential, cf.~Fig.~\mbox{\ref{rho-mean}(c)}.
\par
\begin{figure}%
	\centering
	\includegraphics[width=\linewidth]{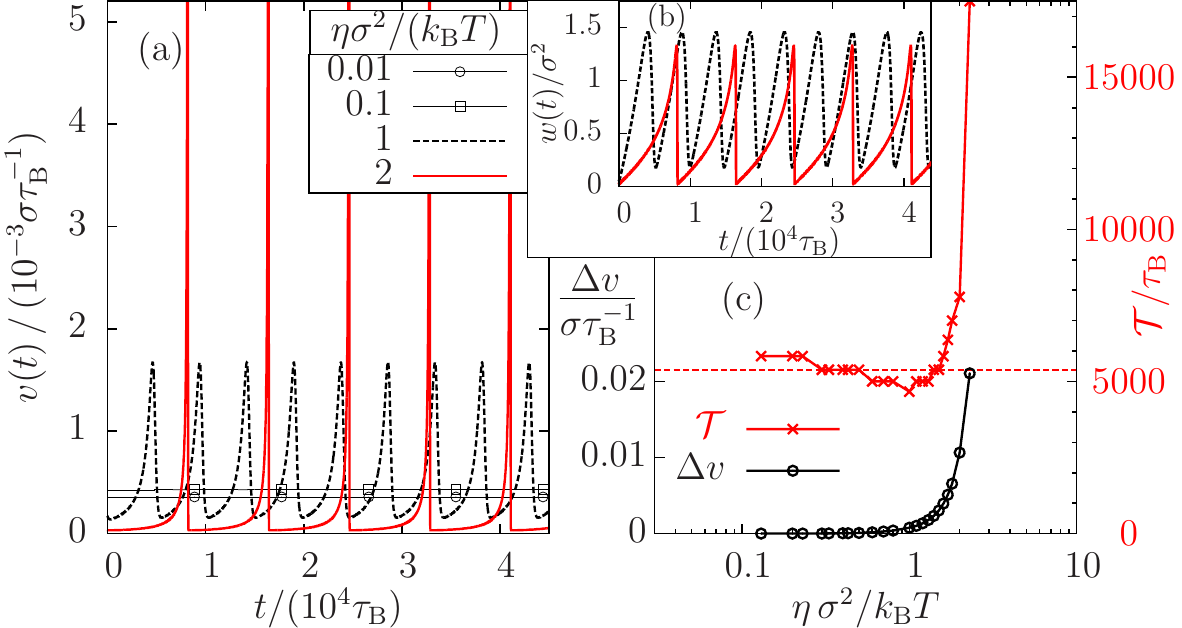}
	\caption{Velocity $v(t)$ (a) and width $w(t)$ (b) as function of time for different confinement strengths $\eta$ for $a=2.5\sigma$. Oscillatory solutions are characterized by sharp peaks. (c) Amplitude $\Delta v$ and period $\mathcal{T}$ of the velocity as function of $\eta$ for $a=2.5\sigma$. Data points are shown only for values of $\eta$ where we find oscillations. The dashed line indicates the time $1/r_K$ where $r_K$ is Kramers' rate.}
	\label{sp-v}
\end{figure}

%
We explain the occurrence of oscillations as follows. We take a view on the beginning of one step of an oscillation at time $t=10^4\tauB$ for $\eta=1\kBT\sigma^{-2}$ [cf.~Fig.~\mbox{\ref{rho-mean}(d)}]. The potential $V_\mathrm{DF}+u$ at this time, shown in Fig.~\mbox{\ref{rho-mean}(c)} as grey shade, shows that the particle is localized at a minimum of $V_\mathrm{DF}+u$.
As time progresses, the constant driving force causes the diffusion of the particle to the next minimum. This leads to a slow increase of the mean particle position $\langle x\rangle$. Then, the feedback control which moves with $\langle x\rangle$, lowers the energy barrier, and steadily accelerates the diffusion through the barrier. This leads to a fast motion until the particle arrives in the next well. The next oscillation then starts again with slow diffusion over the next barrier. The repeated cycle of motion consisting of slow and fast portions is particularly visible in the velocity
\begin{align}
	v(t)&=\frac{d}{dt}\langle x\rangle(t)
	\label{velocity}
	\,,
\end{align}%
and the width $w(t)$ which are plotted in Figs.~\ref{sp-v}(a) and (b), respectively. Notice that $w(t)$ oscillates around a \emph{constant} value, reflecting that the width of the distribution stays finite even at large times (``dynamic freezing'').
\par
\begin{figure}%
	\centering
	\includegraphics[width=\linewidth]{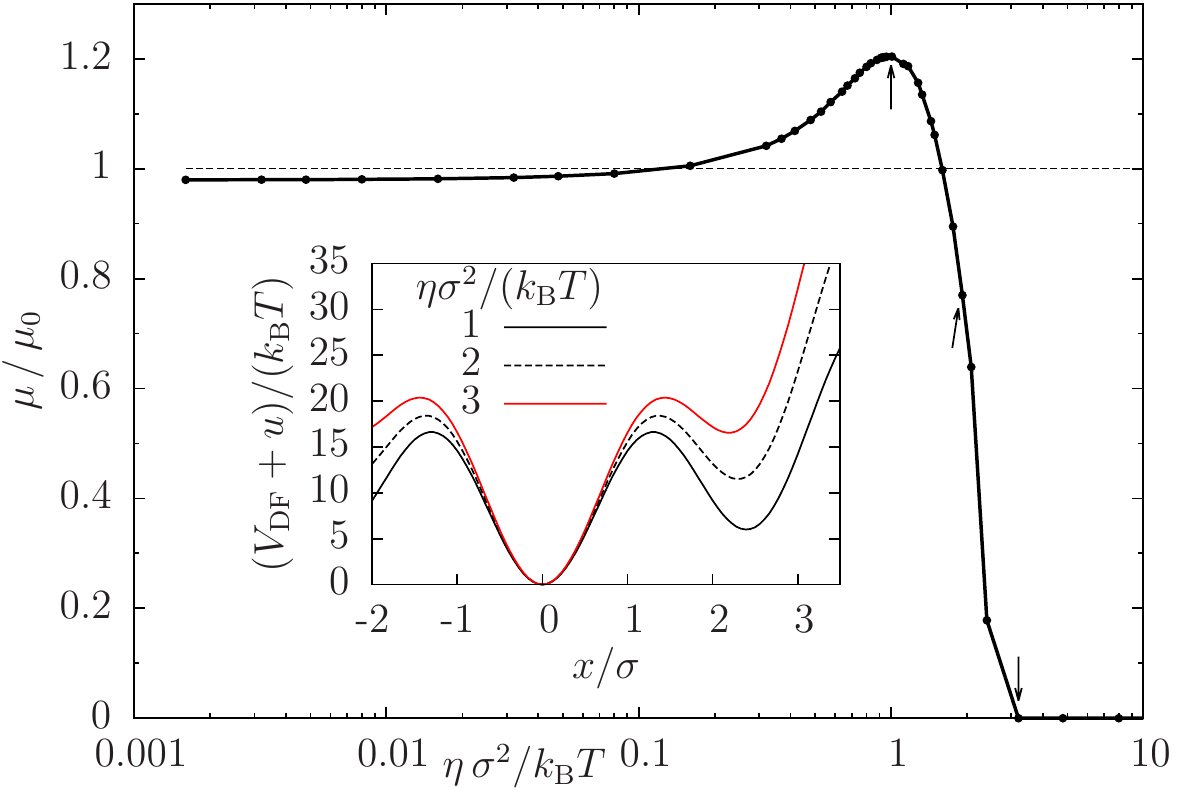}
	\caption{(Color online) Single particle transport: Mobility $\mu$ in dependence on $\eta$ for $a\!=\!2.5\sigma$. The mobility is scaled by the mobility $\mu_0$ of uncontrolled diffusion in a washboard potential. The feedback control can enhance the mobility by up to $\approx 20\%$. Inset: Potential $V_\mathrm{DF}(x,t\!=\!0)+u(x)$ for the three values of $\eta$ which are indicated by crosses in $\mu(\eta)$.}
	\label{sp-mobility}
\end{figure}

%
We analyse the occurrence of these oscillations numerically in terms of period $\mathcal{T}$ and amplitude $\Delta v=(v_\mathrm{max}-v_\mathrm{min})/2$ of velocity, shown in Fig.~\ref{sp-v}(c).
The values $v_\mathrm{max}$ and $v_\mathrm{min}$ are the global maximum and minimum of $v(t)\vert_{t>t_1}$, respectively, where $t_1$ is a time after the disappearance of transients. From Fig.~\ref{sp-v}(c) we find that oscillatory solutions occur in a range of intermediate $\eta$.
In that range the amplitude $\Delta v$ increases with $\eta$ from nearly zero to large values.
Furthermore, the period $\mathcal{T}$ of oscillations roughly coincides with the inverse Kramers rate, which is the relevant time scale for the slow barrier-crossing mentioned before.
As we see in Fig.~\mbox{\ref{rho-mean}}(d), the regime of pronounced oscillations partly coincides with a ``speed up'' of the motion.
We quantify this ``speed up'' via an average mobility based on the time-averaged velocity
\begin{gather}
	\bar{v}=\frac{1}{\mathcal{T}}\int_{t_1}^{t_1+\mathcal{T}}\mathrm{d}t\,v(t)
	\label{meanvelocity}
\end{gather}
such that
\begin{align}
	\mu&=\frac{\bar{v}}{F}
	\label{mobility}
	\,.
\end{align}
Figure \ref{sp-mobility} shows $\mu/\mu_0$ depending on $\eta$, where $\mu_0\approx 1.2\times 10^{-4}/\gamma$ is the mobility of the uncontrolled system ($\eta\!=\!0$) with the same external potential \mbox{\cite{stratonovich58,risken84}}.
For small $\eta$, we find $\mu\approx \mu_0$. The remaining deviation is presumably a numerical issue because, by definition, $\lim_{\eta\to 0}\mu=\mu_0$. At intermediate values of $\eta$ the mobility shows a global maximum which lies above $\mu_0$. From comparison with Fig.~\ref{sp-v}(c) we see that the maximum of $\mu(\eta)$ lies in the range of $\eta$ where the oscillation periods of $v(t)$ are about (in fact, somewhat smaller) than the inverse Kramers rate \cite{risken84,gernert14}. Quantitatively, the maximal enhancement of mobility of $\approx 20.4\%$ is reached at $\eta\approx0.96\kBT\sigma^{-2}$.
For even larger values of $\eta$ a sharp decrease of the mobility to zero is observed -- the motion comes to a halt.
To investigate this phenomenon we first note that the motion is always oscillatory (for these large $\eta$) as long as there is transport at all [compare Figs.~\ref{sp-mobility} and \ref{sp-v}(c)]. From the explanation of the oscillations above, we recall that the oscillation period is determined by the slow diffusion process over the energy barrier. The inset of Fig.~\ref{sp-mobility} shows the potential $V_\mathrm{DF}(x,t\!=\!0)+u(x)$ for three values of $\eta$. To ignite transport the particle must diffuse from the central valley at $x\!=\!0$ to the next valley at $x\approx 2.5\sigma$. The larger $\eta$ the larger the energy barrier. Thus, the larger $\eta$ the smaller the probability that the particle diffuses to the next valley, the longer the period of the oscillations, and the lower the mobility. For $\eta=3\kBT/\sigma^2$ there is no motion at all in the time range of our calculations ($t\le 10^5\tauB$).
\par%
Finally, we note that for single-particle transport the actual value of $a$ is essentially arbitrary because $a$ only determines the scales of time, density, and confinement strength, not the qualitative behaviour.
\subsection{Comparison with open-loop control}\label{sec:openloop}
\begin{figure}%
	\centering
	\includegraphics[width=\linewidth]{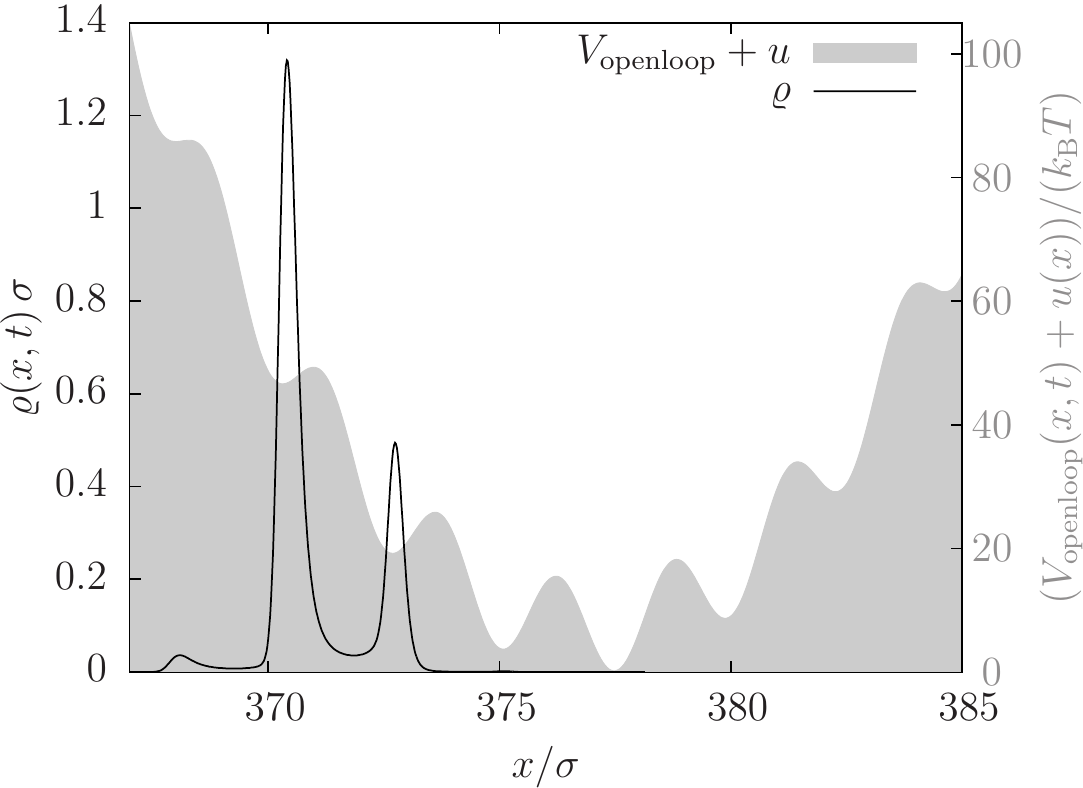}
	\caption{Single-particle transport in presence of the open-loop potential $V_\mathrm{openloop}$ [see Eq.~\eqref{Vopenloop}] with $\eta=1\kBT\sigma^{-2}$ and $v_0= F/\gamma$ at $t=100\tauB$ and $a=2.5\sigma$. Black line: one-particle density $\varrho(x,t)$. Grey shade: potential $V_\mathrm{openloop}+u$.}
	\label{openloop}
\end{figure}

To estimate the benefit of the feedback control scheme over the more established open-loop control, we briefly discuss the motion of a single particle under the potential
\begin{gather}
	V_\mathrm{openloop}(x,t)=\eta\,(x-v_0 t)^2
	\label{Vopenloop}
	\,,
\end{gather}
where $v_0$ is a constant velocity of the trap. Choosing $v_0$ equal to the mean velocity $\bar{v}$ of the feedback controlled system, one observes the same general behaviour, but slight variations of oscillation frequency and amplitude. Large values $v_0>\bar{v}$ lead, by construction, to a fast transport, but the particle is no longer located in the center of the trap. We can see this from Fig.~\mbox{\ref{openloop}} which shows the one-particle density for the velocity $v_0=F/\gamma$ [corresponding to free motion, see Eq.~\eqref{mean-free-diff}] and the effective trap generated by the potential $V_\mathrm{openloop}+u$.
In a real optical trap a large distance of the particle position to the center of the trap implies a large probability to escape \mbox{\cite{balijepalli12,cole12}}. Hence, driving the particle too fast implies the risk of losing the particle completely. On the other hand, being too cautious and driving the particle too slowly is inefficient. Thus, the optimal velocity is difficult to predict in open-loop control. The feedback control automatically finds the optimal driving speed without taking the risk of losing the particle. Furthermore, the feedback control does not influence the direction of motion, it only enhances the absolute value of the mobility.
\section{Many-particle transport}\label{sec:int}
We now turn to interacting systems, as described by the SE \eqref{ddft} with Eq.~\eqref{Fmf} for ultra-soft particles and Eq.~\eqref{Fhc} for hard particles. There are now two relevant length scales, the wavelength of the periodic potential, $a$, \emph{and} the particle diameter $\sigma$. Hence, the wavelength $a$ is not just a scaling factor, as it was the case in single-particle transport. In addition, the number of particles $N$ will play a role because the equations are now non-linear in $\varrho$. In our numerical calculations, we set the ultra-soft particles' interaction strength $\varepsilon$ appearing in Eq.~\mbox{\eqref{Fmf}} to $\varepsilon=4\kBT$. The hard-particle interaction has no parameter besides $\sigma$. The initial condition is set to the equilibrium density resulting at $F=0$.
\par%
In the following we study motion of clusters of interacting particles for various
trap strengths $\eta$, numbers of particles $N$, and dimensionless wavelengths $a/\sigma$.
\subsection{General behaviour}\label{sec:general}
The overall goal is to explore whether particle interactions enhance the efficiency of our feedback control in terms of the mobility.
Before we start with the analysis of the mobility we want to give an impression of the general behaviour of our interacting systems.
\par
\begin{figure}%
	\centering
	\includegraphics[width=\linewidth]{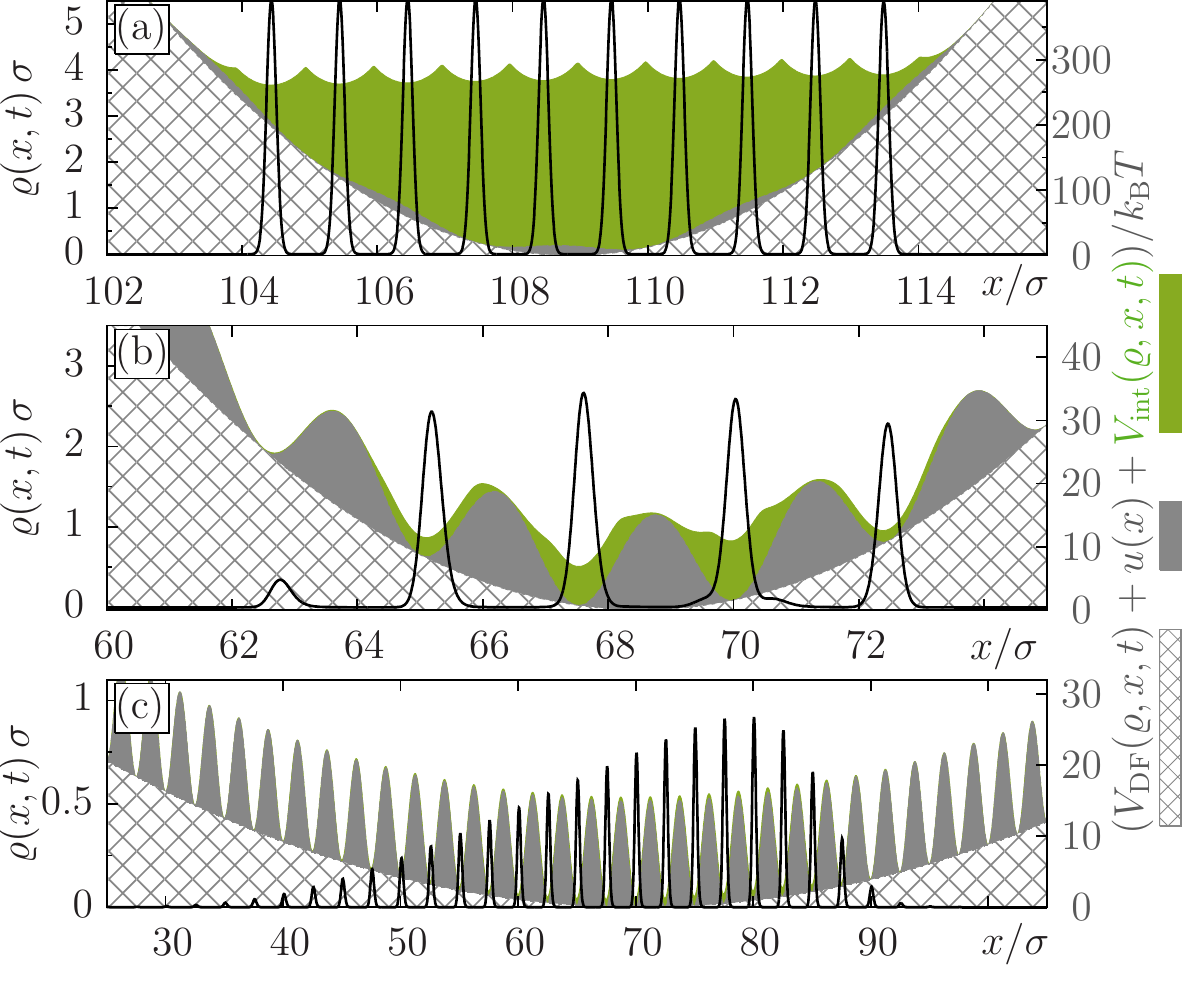}
	\caption{(Color online) One-particle density $\varrho(x,t)$ on the left axis (black lines) and the potentials $V_\mathrm{DF}$, $u$, and $V_\mathrm{int}$ on the right axis. The parameters show (a) the case where $u$ is suppressed, (b) the decrease of barrier height by interaction, and (c) the low-$\eta$/low-$N$ regime. In detail, (a) shows $N\!=\!10$ hard particles forming a chain in a narrow trap ($\eta\!=\!10\kBT\sigma^{-2}, a\!=\!2.5\sigma$) at $t\!=\!28.9\tauB$, (b) $N\!=\!4$ hard particles forming a loose chain in an moderately narrow trap ($\eta\!=\!0.7\kBT\sigma^{-2}, a\!=\!2.5\sigma$) at $t\!=\!985\tauB$, and (c) $N\!=\!4$ ultra-soft particles in a wide trap ($\eta\!=\!10^{-2}\kBT\sigma^{-2}, a\!=\!2.5\sigma$) at $t\!=\!10^5\tauB$.}
	\label{rhopot}
\end{figure}

%
We begin our study of the three-dimensional parameter space ($N,\eta, a$) with small $N$ and small $\eta$. In Fig.~\ref{rhopot}(c) we present a plot of the density profile and the potential $V_\mathrm{DF}+u$ at $\eta\!=\!0.01\kBT\sigma^{-2}, N\!=\!4$.
In fact, the density profiles shown in Fig.~\ref{rhopot}(c) and Fig.~\ref{rho-mean}(a) are very similar. Similarities to the single-particle case vanish gradually if $N$ or $\eta$ (or both) are increased (at constant $a$), yielding larger values of the density in the trap. To describe the effect of these changes in density, we consider the effective potential $V_\mathrm{int}$ that one particle experiences due to the interaction with the other particles. The value of $V_\mathrm{int}$ at a position $x$ increases with the corresponding densities $\varrho(x'), x'\approx x$. Particularly large values of both, $\varrho(x)$ and $V_\mathrm{int}(x)$, occur at the minima of $V_\mathrm{DF}+u$. As a consequence, the potential $V_\mathrm{DF}+u+V_\mathrm{int}$, which governs the motion (together with the constant driving force), is characterized by smaller energy barriers than $V_\mathrm{DF}+u$. Loosely speaking, $V_\mathrm{int}$ fills the valleys of $V_\mathrm{DF}+u$ [see Fig.~\ref{rhopot}(b)]. For high densities, the hard particles form a ``chain'' and the ultra-soft particles form a cluster which is characterised by mutual overlap. In this dense situation the contribution of $V_\mathrm{int}$ to $V_\mathrm{DF}+u+V_\mathrm{int}$ can become so large that $u$ becomes negligible. Thus, there are no hindering energy barriers any more. For both interacting systems, ultra-soft and hard particles, we actually find this case. Fig.~\ref{rhopot}(a) shows the hard particle case as example. The potentials plotted in Fig.~\ref{rhopot}(a) show that $u$ in fact is a minor contribution to $V_\mathrm{DF}+u+V_\mathrm{int}$.
We continue the discussion of parameter variations with focus on the mobility in Sec.~\ref{sec:mobility}.
\par
\begin{figure}%
	\centering
	\includegraphics[width=\linewidth]{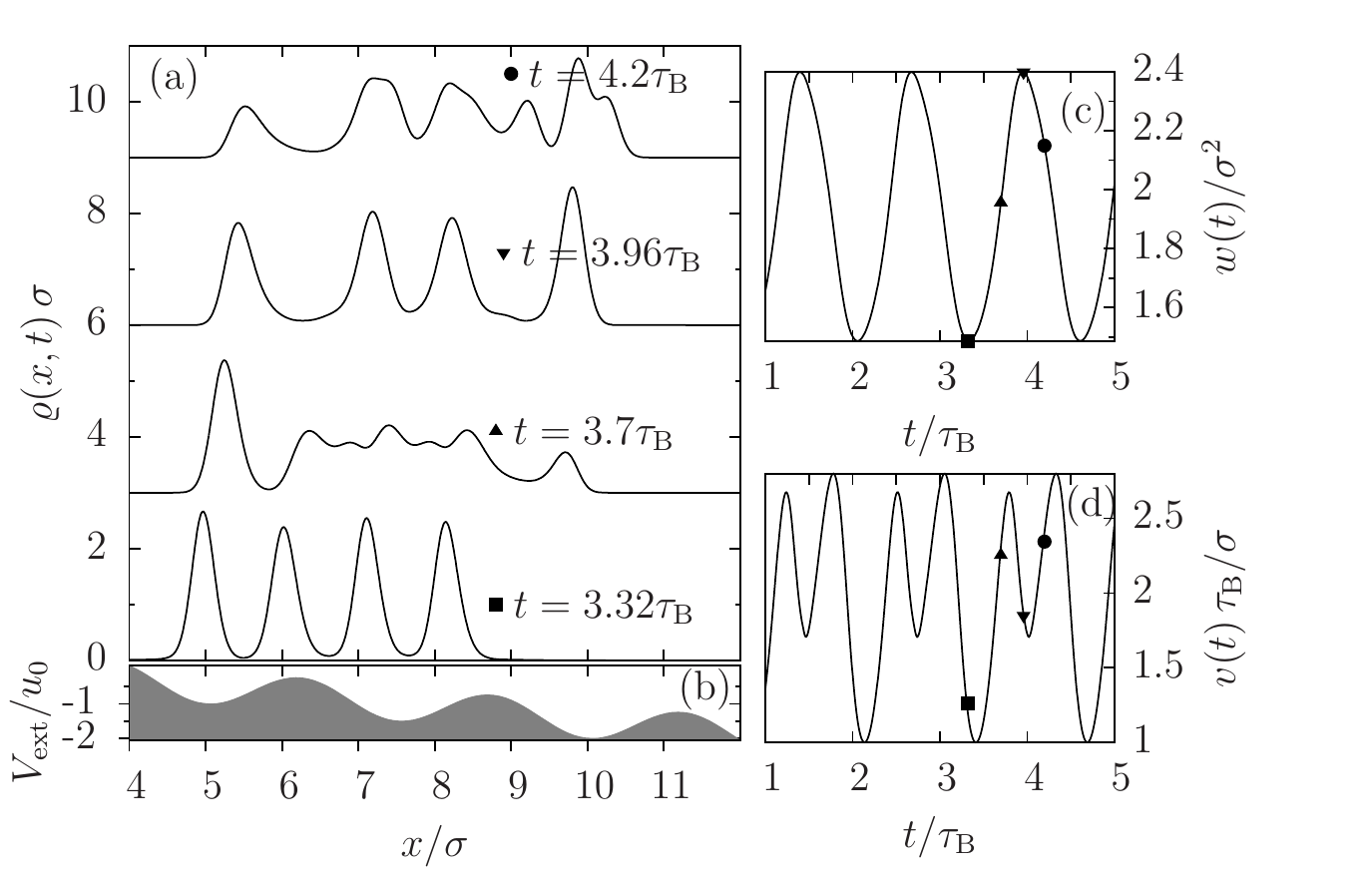}
	\caption{(a) One-particle density $\varrho(x,t)$ for $N=4$ hard particles subject to the tilted washboard potential $V_\mathrm{ext}$ [see part (b)] at $\eta=3\kBT\sigma^{-2}, a=2.5\sigma$ at four times. For clarity the curves are shifted vertically. Each time is labelled with a symbol, reappearing in parts (c) and (d) which show the width $w(t)$, given by Eq.~\eqref{width}, and the velocity $v(t)$, Eq.~\eqref{velocity}, over time, respectively.}
	\label{rhoosc}
\end{figure}

\begin{figure}%
	\centering
	\includegraphics[width=\linewidth]{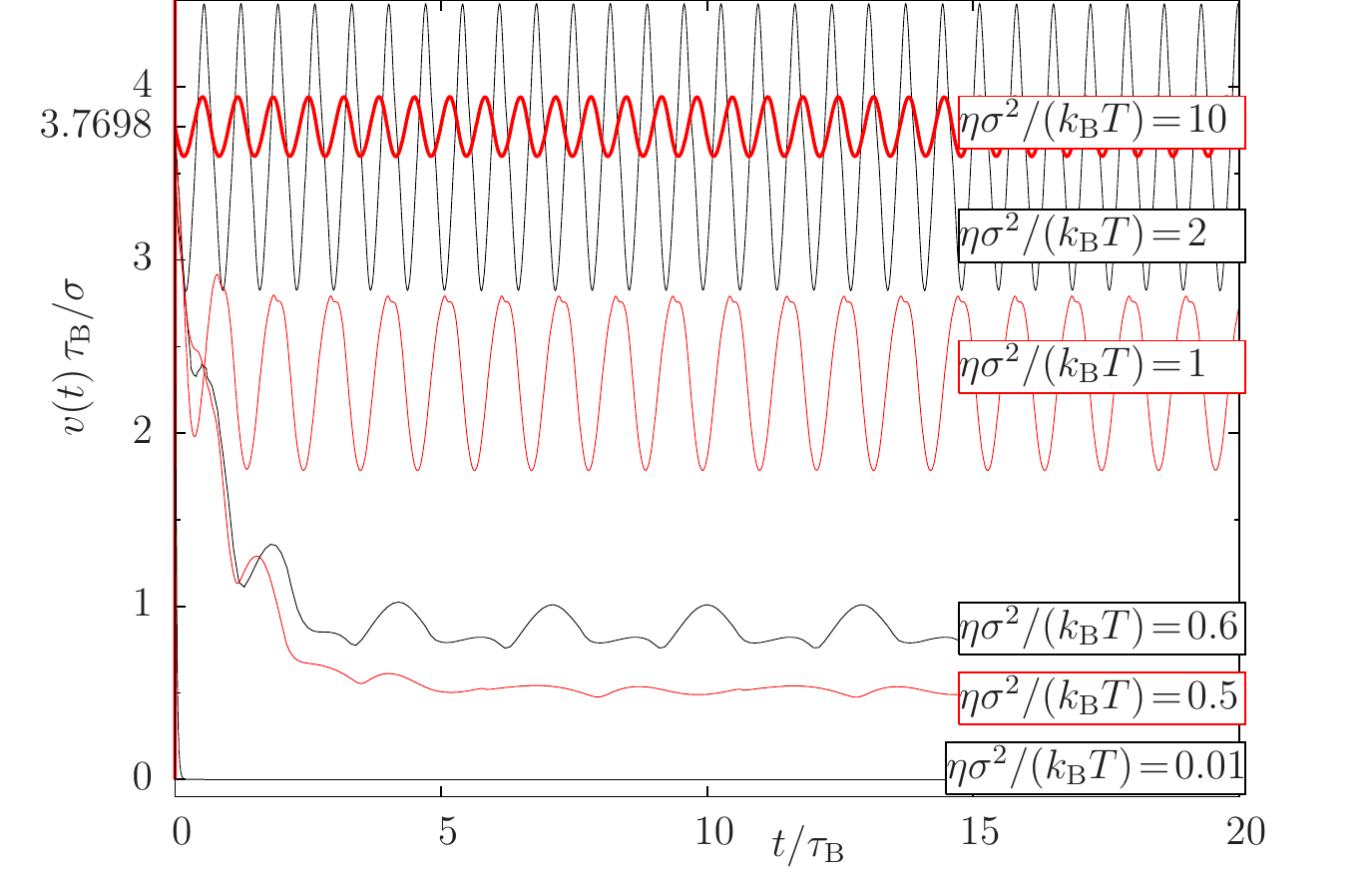}
	\caption{(Color online) Velocity $v(t)$, Eq.~\eqref{velocity}, for $N=10$ hard particles at $a=2.5\sigma$ for a broad range of $\eta$. The velocity $3.7698\sigma/\tauB=F/\gamma$ corresponds to the velocity of free motion under the force $F$. This value is reached at large $\eta$. Oscillations occur for $\eta\ge 0.03\kBT\sigma^{-2}$.}
	\label{velocitytime}
\end{figure}

%
Similar to the single-particle case, we find oscillatory solutions in the range of intermediate to large $\eta$.
For a representative system ($N\!=\!4$ hard particles), Fig.~\ref{rhoosc} summarises different characteristics of the oscillations in terms of width $w(t)$, velocity $v(t)$, and plots of the density for four times during one oscillation period. 
The oscillation period is of the order of $\tauB$ which is much shorter than the oscillation periods of several $10^3\tauB$ we observed in the single-particle case [see Fig.~\ref{sp-v}(b)].
From Fig.~\ref{rhoosc} we see that these oscillations are intimately related to configurational changes while the particle chain moves over a distance of about one wavelength $a$.
Studying $v(t)$ for different $\eta$, see Fig.~\ref{velocitytime}, we find that a couple of different oscillation patterns emerge.
Moreover, the oscillations' frequency rises with the mean of the velocity itself. This can be explained with the observation that the particles move one wavelength $a$ during one period. Note that the maximal amplitude of oscillation neither coincides with largest $\eta$ nor largest mean velocity.
\subsection{Mobility}\label{sec:mobility}
\begin{figure}%
	\centering
	\includegraphics[width=\linewidth]{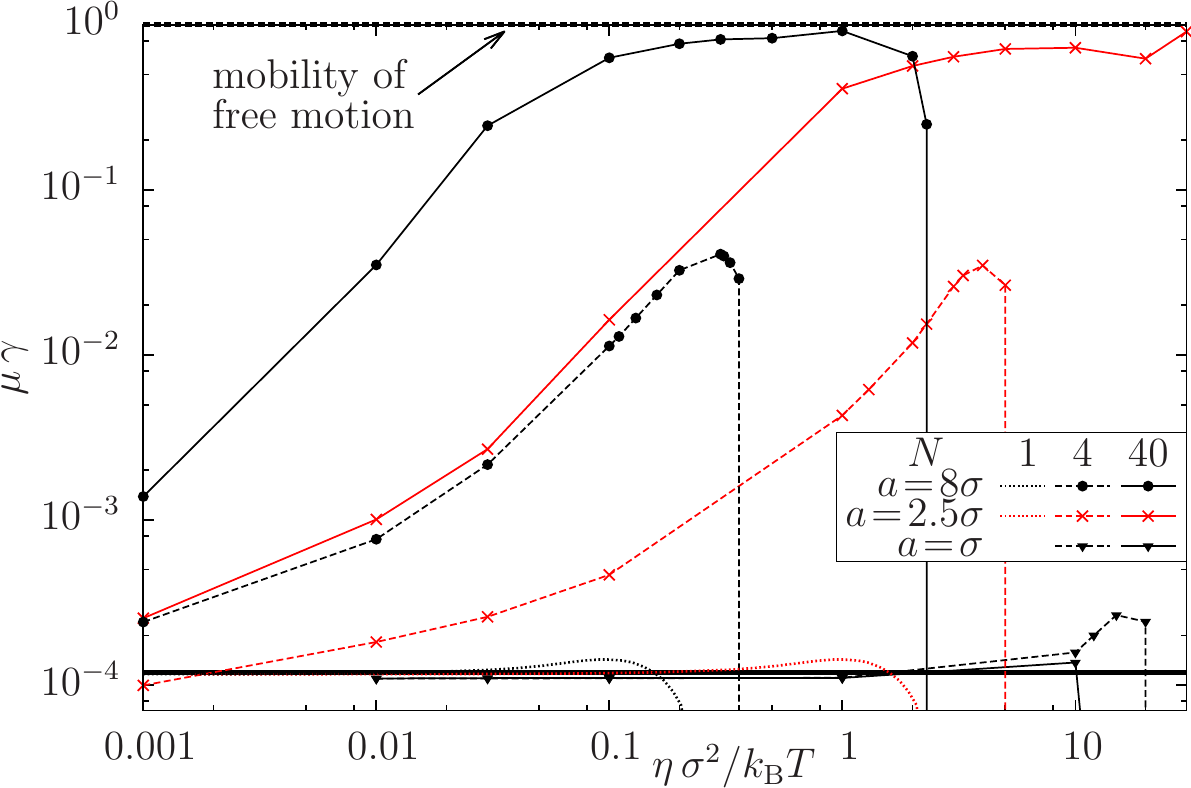}
	\caption{(Color online) Mobility $\mu$ for ultra-soft particles in dependence of $\eta$. Given that enough particles contribute, the mobility can rise to $1/\gamma$, the mobility of free motion. The thick line indicates the mobility in the uncontrolled case. Results from Fig.~\ref{sp-mobility} are included with the notation $N\!=\!1$.}
	\label{mobilitygcm}
\end{figure}

\begin{figure}%
	\centering
	\includegraphics[width=\linewidth]{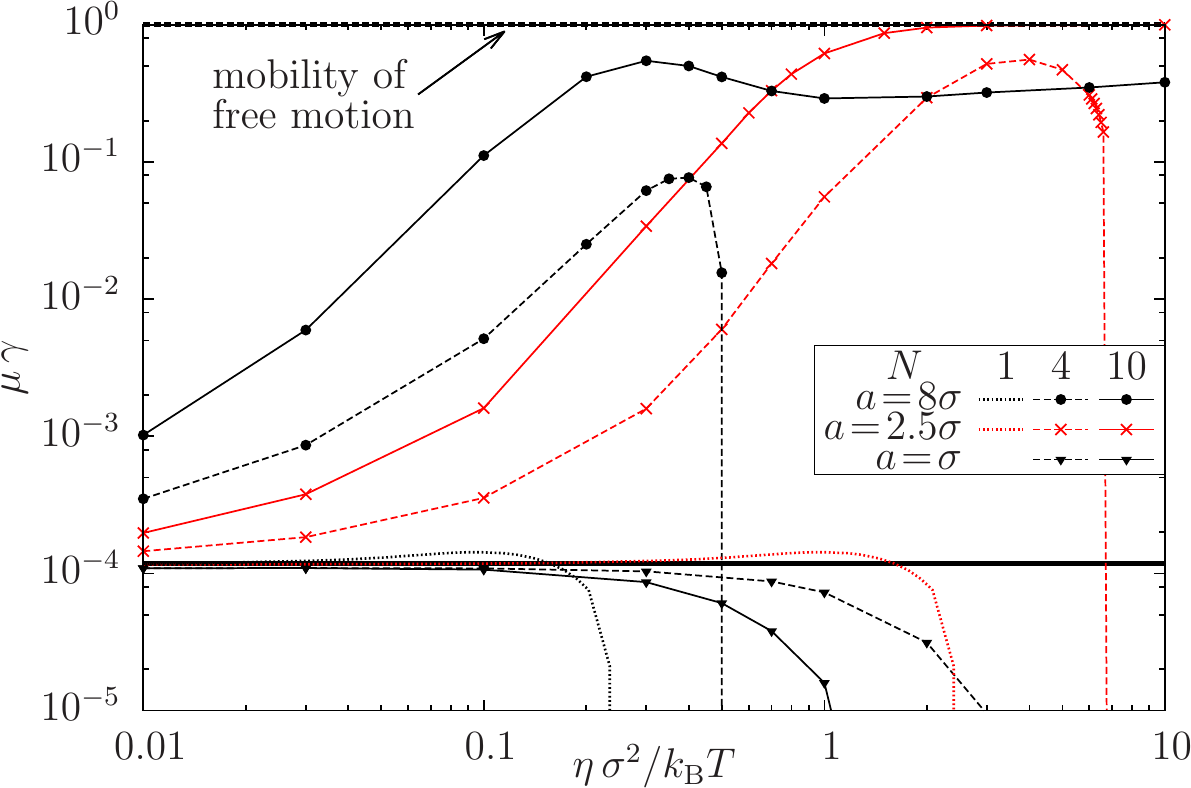}
	\caption{(Color online) Mobility $\mu$ for hard particles in dependence of $\eta$. The mobility reaches $1/\gamma$, the mobility of free motion, if enough particles contribute. The thick line indicates the mobility in the uncontrolled case. Results from Fig.~\ref{sp-mobility} are included with the notation $N\!=\!1$.}
	\label{mobilityhc}
\end{figure}

We now turn to the mobility, as a measure of the efficiency of feedback control. We define the mobility $\mu$ in the same way as in the single-particle case via Eq.~\eqref{mobility}. Figures \ref{mobilitygcm} and \ref{mobilityhc} show $\mu$ in dependence of $\eta$ for ultra-soft and hard particles, respectively.
\par%
For $a>\sigma$ we observe an extreme growth of $\mu$ with $\eta$ and $N$ over several orders of magnitude for both particle species. We explain this behaviour with the corresponding decrease of the height of the energy barriers in $V_\mathrm{DF}+u+V_\mathrm{int}$, which results in a larger diffusion rate and a faster transport. The same effect was observed in a study of the transport of super-paramagnetic colloids \cite{kreuter13}. For certain $\eta$ and $N$, $\mu$ increases even up to the maximal possible value $\mu=1/\gamma$, the mobility of free motion. An example for this large mobility is the case of $N\!=\!10$ hard particles at $\eta=10\kBT\sigma^{-2}$, shown in Fig.~\ref{rhopot}(a). In this case, there are no hindering energy barriers (as we have analyzed in Sec.~\ref{sec:general}) which then results in the high mobility.
To achieve this suppression of $u$ the well created by the trap potential must be very deep, i.e.~$300\kBT$ for the exemplary case shown in Fig.~\ref{rhopot}(a). This value exceeds those in typical experiments with light fields \cite{juniper12,lopez08}. However, the transport of $N\!=\!40$ ultra-soft particles at the mobility $\mu=1/\gamma$ at $a=8\sigma$ needs a trap which is only $40\kBT$ deep.
\par%
Further, we see from Figs.~\ref{mobilitygcm}, \ref{mobilityhc} that our feedback control does not lead to a significant speed up for $a\!=\!\sigma$. By analysing the potential landscape for $a=\sigma$ for a series of $\eta$ and $N$ (not shown) we find that the effective potential $V_\mathrm{int}$ develops peaks \emph{between} the minima of $V_\mathrm{DF}+u$. This means that the effective potential barrier encountered by a moving particle \emph{increases} when $\eta$ or $N$ is enlarged. This is in contrast to the case $a> \sigma$ where the peaks of $V_\mathrm{int}$ are found \emph{at} the minima of $V_\mathrm{DF}+u$ [see Fig.~\ref{rhopot}(b)]. Our interpretation for the case $a=\sigma$ therefore is that the particles ``pin'' each other to the potential minima of $u(x)$. 
\par%
We now consider the behaviour of $\mu(\eta)$ for large $\eta$ (Figs.~\ref{mobilitygcm}, \ref{mobilityhc}). For small $N$ we observe a breakdown of motion, similar to the one observed in the single-particle case (see Fig.~\ref{sp-mobility}). However, this breakdown is shifted towards larger values of $\eta$. We recall that an increase of $N$ at fixed $\eta$ (Sec.~\ref{sec:general}) leads to a decrease of the barriers of the potential $V_\mathrm{DF}+u+V_\mathrm{int}$. This enhances the mobility (relative to that at $N\!=\!1$) in the first place. However, upon increase of $\eta$ (at fixed $N$) there can be a situation where the diffusion rate is not sufficient any more to populate the next local minimum of the potential $V_\mathrm{DF}+u$. This is where transport breaks down. The combination of these two effects leads to the observed shift of the breakdown of mobility. Upon further increase of $N$ and $\eta$, there comes a point where the large energy scales of $V_\mathrm{DF}$ and $V_\mathrm{int}$ suppress any influence from $u$. Therefore, we expect that the transport for high $N$ exists for arbitrarily large $\eta$.

In Fig.~\ref{mobilityhc} we see that the increase of $a$ at constant $\eta$ and $N$ leads to an enhancement of mobility (as long as there is transport at all). This can be explained with the potential $V_\mathrm{DF}+u$, whose valleys become broader the larger $a$. In a broader valley more particles accumulate which strengthens the role of interaction for the barrier crossing. However, this effect is limited by $N$: The particle number must be large enough to fill at least one valley with particles, otherwise the transport breaks down.
\subsection{Time delay}
\begin{figure}
	\centering
	\includegraphics[width=\linewidth]{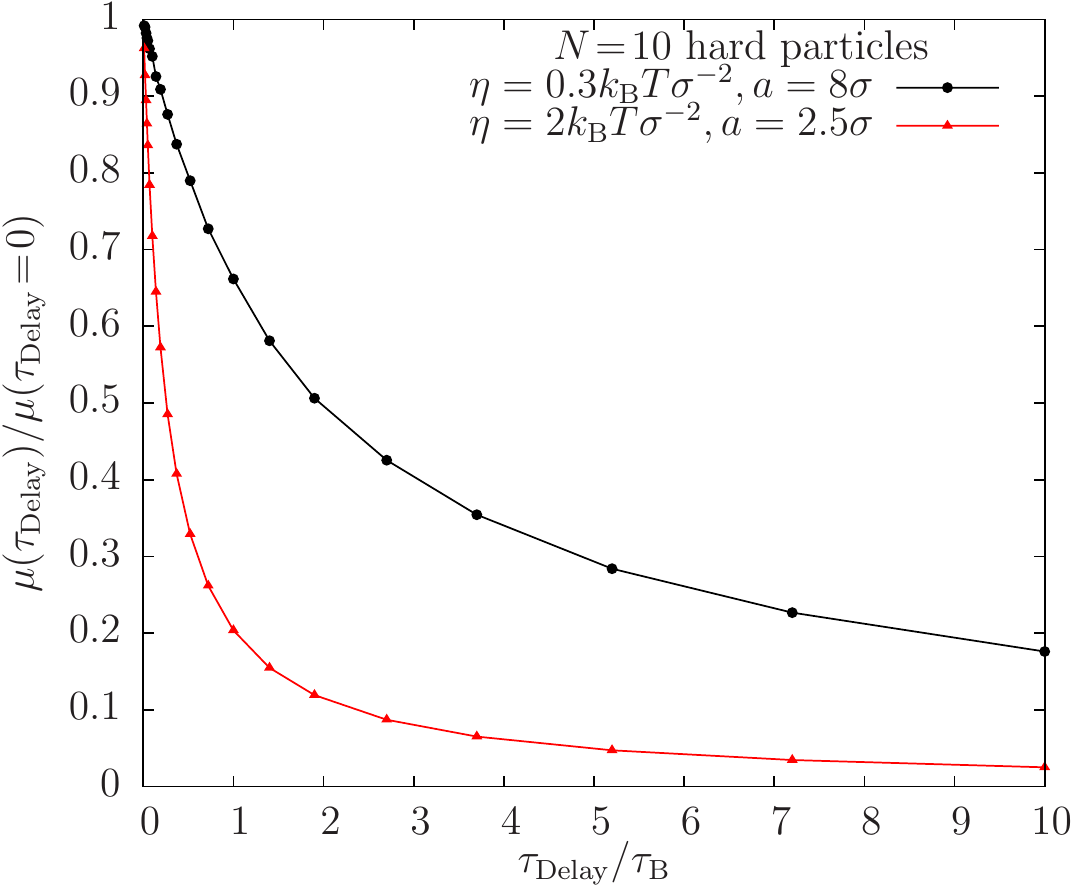}
	\caption{Effect of time delay on mobility $\mu$ for hard particles. Even for very long delays the mobility is only reduced by one order of magnitude.}
	\label{delay}
\end{figure}

In a realistic set up with feedback control, a finite time is required to perform the measurement required to define the control (In the present case, this measurement process concerns the average particle position). Hence, there is a certain \emph{time delay} $\tau_\mathrm{Delay}$. To explore the sensitivity of our results towards $\tau_\mathrm{Delay}$ we change the control potential given in Eq.~\eqref{Vdf} into the expression
\begin{align}
	V_\mathrm{DF}^\mathrm{delay}(x,\varrho)=\eta\,(x-\langle x\rangle(t-\tau_\mathrm{Delay}))^2
	\label{Vdfdelay}
	\,.
\end{align}
We now consider two special cases involving hard particles, where the non-delayed feedback control leads to a particularly high mobility (see Fig.~\ref{mobilityhc}). Numerical results are shown in Fig.~\ref{delay}. The delay causes a pronounced decrease of mobility which appears to be linear in $\tau_\mathrm{Delay}$ for small delay times.
Realistically, feedback mechanisms can be implemented at the time scale of $10$\,ms \cite{lopez08,cohen06,bregulla14} whereas $\tauB$, the timescale of Brownian motion, is for $\mu$m-sized particles in the order of $300$\,ms \cite{lopez08} or larger \cite{dalle11,lee06}. Hence, we expect that the ratio $\tau_\mathrm{Delay}/\tauB$ is rather small, that is, of the order $10^{-1}$. For such situations, our results in Fig.~\ref{delay} predict only a small decrease of $\mu$ relative to the non-delayed case. However, even for large delays the mobility only decreases about one order of magnitude. This implies that even the time delayed feedback control can enhance the mobility by more than two orders of magnitude with respect to the uncontrolled case.
\section{Conclusion}\label{sec:conclusion}
Inspired by the physics of a moving optical tweezer, we have proposed a feedback control strategy for the 
collective transport of interacting colloids through a corrugated potential landscape. 
Our main goal was the theoretical demonstration of the working principle
for a well-defined model system. 
To this end we have considered the one-dimensional, overdamped motion of colloids with either hard or soft repulsive interactions in a tilted washboard potential. 
The feedback control enters into the (Smoluchowski-like) equation of motion via 
a harmonic potential centered at the mean particle position. Thus, contrary to other studies \cite{abreu11,bauer12}, the present feedback control cannot induce motion on its own.

The main result of our study is that the interplay of the feedback control, on one hand, and particle interactions, on the other hand, can generate a drastic increase of the average mobility
by several orders of magnitude relative to the uncontrolled, single-particle reference case. The largest mobilities occur for rather stiff traps and high densities (i.e., large $N$) inside the trap,
yielding chain- or cluster like packages of colloids. Here, the mobility rises up to its limiting value defined by the mobility of a freely moving, overdamped particle. Interestingly, this giant increase does not occur for a single particle under the same feedback control. This shows that the observed mobility enhancement is indeed an interaction effect. The enhancement can be explained by the fact that, in presence of particle interactions, these dominate the effective ``field" acting on an individual particle, while the impact of the external potential barriers vanishes. Thus, particles ``help each other" to overcome the external barriers. Another new feature is the emergence of {\em oscillatory} behavior of the mean velocity 
(and the width of the density distribution) due to the feedback-controlled trap. The latter effect occurs for both, single and interacting particles, with the period of oscillations being close to the inverse of Kramers' escape rate. 

From an application point of view it is interesting that, due to its coupling to the mean position, the feedback-controlled trap implies a small risk to ``lose'' particles. Indeed,
the width of the distribution stays constant on time-average, reflecting a ``dynamical freezing". This is different from externally moved, ``open-loop" traps, where an inappropriate choice of the trap velocity easily lead to a broadening of the density distribution, and thus, a spreading of particles out of the trap (see discussion in Sec \ref{sec:openloop}). Another experimentally relevant issue concerns the impact of time delay(s). Here we have shown that time delay does indeed reduce the mobility, similar to what has been observed
in ratchet systems \cite{craig08-024}. However, for realistic time delays the remaining mobility is still enhanced by two orders of magnitude.

Concerning the methodology, we note that the DDFT scheme employed here implies an ``adiabatic" approximation of the time-dependent two-particle correlations. It is now well established \cite{marconi99,fortini14,reinhardt12} that this approximation may generate artefacts especially for densely packed particles, e.g., during the expansion of a cluster. Since we are mainly focusing on steady transport conditions we expect our results to be at least qualitatively right. Still, it would be very interesting and important to test our predictions against explicit Brownian Dynamic simulations of the corresponding (overdamped) Langevin equation, Eq.~\eqref{LE}.

Finally, we would like to point out that the concept behind dynamical freezing is not restricted to one-dimensional washboard potentials. Indeed, the present feedback control can easily be formulated in two or three spatial dimensions. Further, the external potential hindering the motion does not have to be static or even periodic which enriches possible applications. An interesting question is how well the present control strategy works for other types of colloidal interactions, particularly attractive ones. Another open question concerns the implications for the (non-equilibrium) thermodynamics of the system, an area which currently receives much attention \cite{loos14,seifert12,sagawa13,munakata14}. Work in these directions is in progress.
\section{Acknowledgement}
We gratefully acknowledge stimulating discussions with C.~Emary. This work was supported by the Deutsche Forschungsgemeinschaft through SFB-910. 

\end{document}